# The Recurrent Processing Unit: Hardware for High Speed Machine Learning

Heidi Komkov[†], Alessandro Restelli, Brian Hunt, Liam Shaughnessy, Itamar Shani, Daniel P. Lathrop
University of Maryland
College Park, USA
[†]heidib@umd.edu

*Abstract*—Machine learning applications are computationally demanding and power intensive. Hardware acceleration of these software tools is a natural step being explored using various technologies. A recurrent processing unit (RPU) is fast and power-efficient hardware for machine learning under development at the University of Maryland. It is comprised of a recurrent neural network and a trainable output vector as a hardware implementation of a reservoir computer. The reservoir is currently realized on both Xilinx 7-series and Ultrascale+ ZYNQ SoCs using an autonomous Boolean network for processing and a Python-based software API. The RPU is capable of classifying up to 40M MNIST images per second with the reservoir consuming under 261mW of power. Using an array of 2048 unclocked gates with roughly 100pS transition times, we achieve about 20 TOPS and 75 TOPS/W.

## I. INTRODUCTION

Spurred by the continual advancement of CMOS fabrication technology, digital computers have attained unprecedented computational power and efficient data storage. While the foundations of neural networks were established in the last century, machine learning has only exploded in popularity in recent years due to the availability of fast and inexpensive computing hardware. Currently, machine learning applications are in high demand for edge computing applications—for example on wireless devices with strict power and memory budgets, or in scenarios with poor network connectivity or stringent security requirements where data cannot be sent off the device for processing. At the same time, the end of Moore's law is on the horizon due to fundamental physics limitations, meaning that advancements in computing capabilities may slow down. To continue the pace of advancement of computing hardware, and to develop new approaches where existing algorithms fall short, non-Von Neumann architectures must be considered. In this paper, we present a new hardware design for a specific class of neural networks known as reservoir computers, which can be implemented in ASICs or commercial off-the-shelf FPGAs [1].

## II. RESERVOIR COMPUTING USING AUTONOMOUS GATE ARRAYS

### A. Reservoir Computing

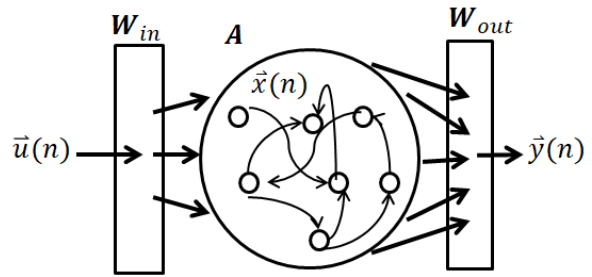

Fig. 1: The reservoir computer is comprised of input matrix $W_{in}$, which determines the mapping of input vector $\vec{u}(n)$ to the reservoir; a sparse, randomly connected reservoir of nonlinear nodes with adjacency matrix $A$; and output vector $W_{out}$ which maps the reservoir state to the desired output $\vec{y}(n)$.

Reservoir computing is a biologically inspired machine learning technique developed in the early 2000s [2-3]. The reservoir computer consists of an input layer, a recurrent neural network termed the reservoir, and an output layer. Since it is difficult to train the weights in a recurrent neural network through backpropagation due to issues with vanishing or exploding gradients, the reservoir computer leaves the input weight matrix $W_{in}$ and the reservoir adjacency matrix $A$ fixed after their random initialization. Training computes only the output weight matrix $W_{out}$, which is used to map the states of nodes in the reservoir to the desired output. During training, the weights of $W_{out}$ are adjusted with linear or logistic regression, and during test, the output is computed with a single matrix multiplication. The reservoir is allowed to evolve autonomously for some time before the output is computed.

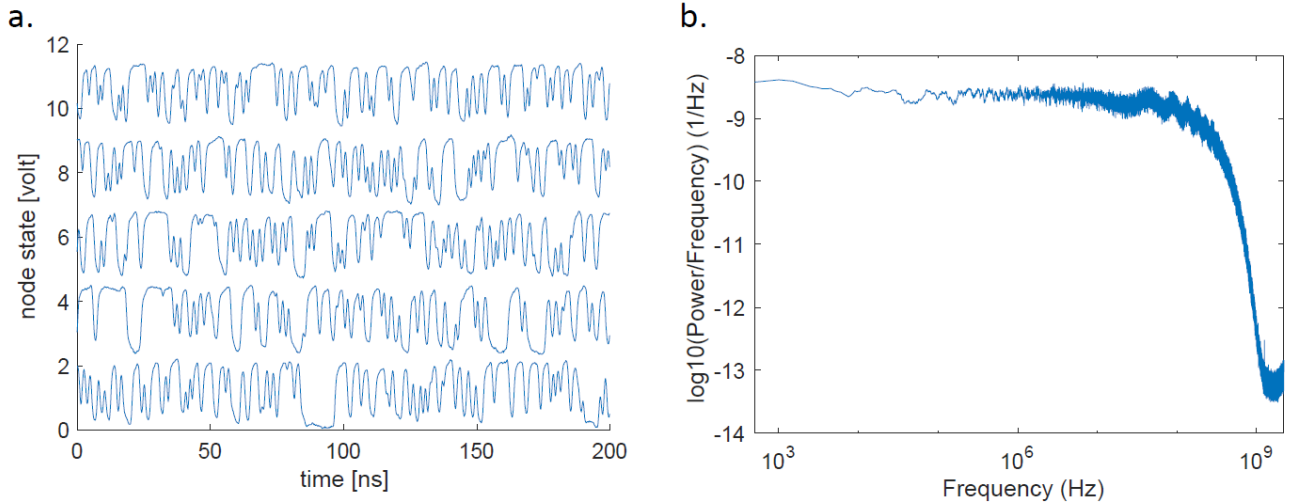

Fig. 2: Dynamics of single nodes in a 512-node all-XOR network. On the left, time traces from directly probed pins on an Altera Cyclone III FPGA are shown, with different traces having a vertical offset. On the right, the frequency spectrum of the signal showing broadband noise is plotted. This amount of self-excitation in a network is excessive for a reservoir, but can be controlled by varying the ratios of gates to change mean sensitivity as we detail here. Reproduced from [4].

The reservoir is effectively transforming the input vector and its recent history to a higher dimensional space. The output layer is trained to decide which combination of those transformations produces the desired result. Memory in the reservoir comes from its recurrent nature, which causes information to circulate before fading out. The mixing of information that was introduced to the reservoir at different times makes the reservoir particularly suited to predicting time-series data, because the reservoir has the ability to form associations between successive points in time. However, the reservoir computer can also be used on one-time-step inference problems.

Perhaps the simplest way to instantiate a reservoir is in software. When simulated in software, the behavior of the reservoir can be described using the discrete-time equations below.

$$\vec{x}(n) = f_{NL}\big(\boldsymbol{A}\vec{x}(n-1) + \boldsymbol{W}_{in}\vec{u}(n)\big)$$
$$\vec{y}(n) = \boldsymbol{W}_{out}\vec{x}(n)$$

Here $\vec{x}(n)$ is the state vector of the reservoir nodes, and $\vec{u}(n)$ is the input data vector at time step $n$. Each node applies the nonlinear activation function $f_{NL}$, which is typically a hyperbolic tangent function. The adjacency matrix $\boldsymbol{A}$ has real weights so that $A_{ij}$ is the strength and sign of the connection within the reservoir from $x_j$ to $x_i$. Similarly, $\boldsymbol{W}_{in}$ contains the input weights from $\vec{u}(n)$ to $\vec{x}(n)$. Optionally, an additional feedback connection can be represented by a matrix $W_{fb}$ to couple the output to the input at the next time step, which can be useful for time series prediction tasks [5].

Software models of reservoir computers have been successfully used for chaotic time series prediction, speech recognition [6], and early seizure detection [7], and dynamic system control [8] among many other applications [9]. However, a reservoir realized from a physical nonlinear dynamical system, as opposed to a simulation, has the advantage that it may run at lower power and at higher speed than on a conventional computer. To this end, inherently nonlinear photonic, electronic, and microelectromechanical systems have been implemented as reservoirs [9 -11].

### B. Autonomous gate arrays

To take advantage of the low cost and scalability of conventional CMOS processes, we use an autonomous Boolean gate network as the reservoir. An autonomous gate array is a network of Boolean logic gates interconnected by wires and running unclocked, in an analog fashion, at the speed determined by the propagation delays through the gates. In contrast to a simulation, the gates are inherently parallel. Studies of Boolean networks have revealed their varied dynamics ranging from the periodic behavior of ring oscillators to chaotic behavior being used for random number generation using behavior of a self-excited network as is shown in Fig. 2 [11-12]. While our current reservoir computer studies are performed on an FPGA due to the flexibility of reconfiguring such a platform, the same designs can later be implemented on a dedicated chip (an ASIC) for further improvements in speed and power reduction. The gates in the FPGA are unclocked, but in order to interface with external synchronous data systems, the readouts are clocked, which limits how accurately the internal dynamics can be probed.

We randomly choose, with some rules, the gate types, input-to-gate, and gate-to-gate connections. The sparsity of connections is similar to the use of sparse matrices $\boldsymbol{A}$ and $\boldsymbol{W}_{in}$ in a software implementation. Although the Boolean design does not allow variable connection strengths, the use of varying gate types adds useful heterogeneity. In the next section, we illustrate how we can adjust the network dynamics by varying the composition of gate types.

### C. Network Dynamics

The dynamics of the network determine the utility of the reservoir for inference. An overly excited network such as the one shown in Fig. 2 exhibits turbulent self-excitation even in the absence of stimulation, which prevents a reproducible response to input data. Time traces of an all-XOR network of 1120 nodes are shown in Fig. 2a, and the broadband chaos that the nodes exhibit is in Fig. 2b. On the other hand, networks exhibit useful computational properties when they have transient activity that fades out in finite time after an input stimulus. Fig. 3 is an example of the collective response of a network which exhibits no self-excitation, unlike the all-XOR network, but which has a long transient response. The plot shows mean distance and standard deviation of the network from its steady state, defined as follows:

$$d_t = \langle |M_{t,p,n} - m_n| \rangle_{p,n}$$
$$s_t = \langle \sigma(M_{t,p,n})_p \rangle_n$$

Here $m_n$ is the steady-state value of node $n$ and $M_{t,p,n}$ is the state of node $n$ at time $t$ after pulse $p$. The distance $d_t$ indicates how the network state has departed from its steady state, averaged over many pulses. The standard deviation between pulses, $s_t$, indicates the degree to which there is discrepancy between network states after identical pulses. These curves show that the network's response is partially, but not completely repeatable. However, networks of this type can successfully be used as reservoirs.

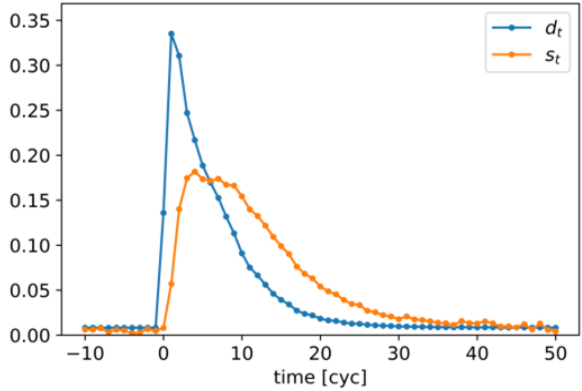

Fig. 3: Temporal properties of an autonomous gate array that is used for computation on the PYNQ-Z1. Of 1120 nodes, statistics were taken on 80% of the 1120 nodes of this network with $\bar{S} = 0.504$. This is a network which does not show self-activity, but exhibits a long transient. Reproduced from [4].

The overall activity level of a network is strongly influenced by the sensitivity of the Boolean logic gates that comprise it. For a single gate, we define sensitivity ($S$) as the fraction of changes in input which result in a change of the output. Fig. 4 below shows graphs of transitions of XOR and OR gates. The color of vertex represents the output of the logical function given the input indicated next to it. Edges are drawn between results when their inputs have a difference of only one bit. For an XOR, because the output changes for every change in input, all edges are counted, and the XOR gate has the maximum sensitivity $S = 1$. By the same reasoning, the OR gate has $S = 0.5$.

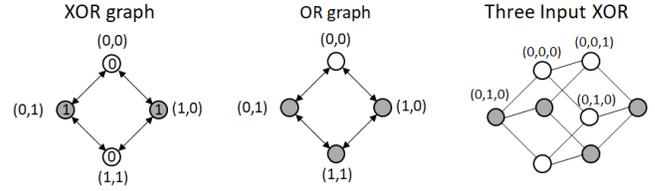

Fig. 4: Transition graphs of XOR and OR gates aid in calculating gate sensitivity. Vertices are colored according to the Boolean logic function output. Edges are drawn between vertices whose inputs vary by one bit. Sensitivity is the fraction of edges drawn between vertices of different colors.

Average network sensitivity has a large influence on the activity level of network with constant input in time. Fig. 5 demonstrates this effect for a variety of networks of varying average sensitivity, in the absense of any external input. The activity level $C$ of the network is defined as the L1 average rate of change of node states:

$$C = \langle |x_i(n) - x_i(n-1)| \rangle_{i,n}$$

A mean sensitivity $\bar{S}$ of less than 0.5 is desirable for reproducibility of network response to the same input. Values of $\bar{S}$ close to 0.5, as in Figure 3, typically have longer memory than smaller values of $\bar{S}$, which may be advantageous for some applications.

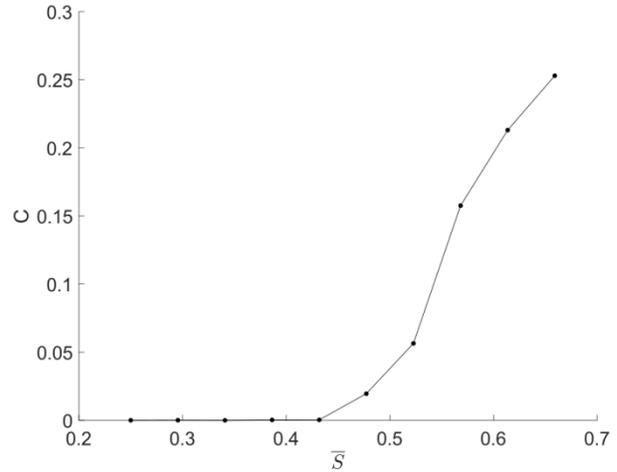

Fig. 5: Network activity level as a function of average gate sensitivity $\bar{S}$. The average is taken over four randomly chosen networks with the same sensitivity.

### III. BENCHMARKING

#### A. Reservoir Processing Unit Hardware

The reservoir used in these tests is implemented in a PYNQ-Z1 board based on the system on chip (SoC) ZYNQ XC7Z020-1CLG400C from Xilinx® [14]. The SoC features a dual-core Cortex-A9 processor as processing system (PS) and Artix-7 family programmable logic (PL). The PS runs PYNQ (Python Productivity for Zynq) an open source framework that allows to accelerate certain system design steps using Python. Specifically the PS, programmed in Python, manages all communication through Gigabit Ethernet and all control tasks and reads and writes data to random access memory (RAM) buffers in the PL. The PL has one input RAM buffer that is 1024 cells deep and 1024 bits wide and an output buffer that is 1024 cells deep and 2048 bits wide. A synchronous state machine clocked at 200 MHz reads the content of the 1024x1024 bits input RAM every clock cycle and applies the 1024 bit pattern to the input of the reservoir while at the same time the 2048 outputs of the reservoir are sampled and stored in the 1024x2048 bits output RAM buffer. The reservoir is specified in a Verilog file as pure combinatorial logic with the gate types and the connectivity we described in II-B. Each node (i.e. gate) of the network is synthesized at the register transfer logic (RTL) level within a single look-up table of the FPGA. Specific directives are inserted in the code for retaining most gate-to-gate nets that would otherwise be simplified out by the RTL synthesizer. The design has also been ported to a ZCU104 Xilinx® evaluation board, based on the Zynq UltraScale+ XCZU7EV-2FFVC1156. The waveforms shown in Fig 2 are captured from an Altera Cyclone III evaluation board in which only the reservoir is synthesized, and a small fraction of the reservoir output nodes are mapped to FPGA outputs, allowing them to be read directly by an oscilloscope.

To implement a reset function, each gate is connected to a common reset signal and the OR operator is applied between the reset and the result of the gate logical operation to produce the actual gate output. In this way when the reset signal is asserted all gate output nets will be in logic state 1 and this produces a stable and reproducible initial state for the whole network independent from the current input.

#### B. MNIST image classification

The MNIST dataset is a collection of greyscale handwritten digits from 0 to 9. Its 60,000 training and 10,000 test images are commonly used as an initial benchmark for image classification algorithms. Fig. 6 illustrates how data is presented to the network. To ensure consistency, the reservoir is reset to bring it to a consistent steady state. The $28 \times 28$ pixel MNIST image is thresholded at 34% of maximum image brightness to binarize it, then rearranged into a single column vector, which is introduced to nodes in the reservoir all at once. Several clock cycles later, the complete reservoir state is read out at 200MHz and transmitted to a supervisory computer, where the output layer is computed by minimizing either the mean squared error or the cross-entropy cost function used in logistic regression. The reservoir can process up to 40M images per second, with individual node transition times on the order of 100ps, making the reservoir speed 20 TOPS. On the PYNQ, with a reservoir of 2048 nodes, the power consumption of the board is measured to rise by 261mW when the reservoir is activated. The overall board power is 6W.

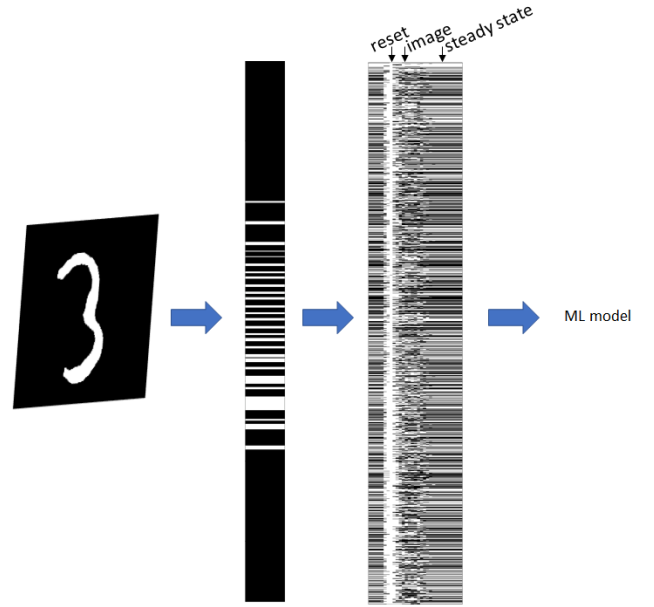

Fig. 6: How the RPU processes MNIST image data. The 28x28 pixel image is unraveled into a column vector. It is presented to the reservoir after a reset. Reproduced from [4].

TABLE I.    RPU ACCURACY IMPROVEMENT OVER LINEAR CLASSIFIER FOR MNIST CLASSIFICATION

|  | Linear regression | Logistic regression |
|---|---|---|
| **Linear classifier only** | 14.0% error | 9.0% error |
| **RPU with linear classifier** | 12.4% error | 7.0% error |

### CONCLUSION

The RPU is a reservoir computer that can operate at high speeds and at low power, designed for implementation in CMOS. We show that the RPU offers a significant reduction in error rate over linear classifiers. As well, our design shows outstanding speed and low-power edge computing capabilities. Furthermore, beyond image analysis demonstrated here, the short-term memory property of certain reservoirs illustrated in Section II.C can be demonstrated to be advantageous in the processing of time series signals including radio signal frequency classification and audio signals.

### ACKNOWLEDGMENT

This material is based upon work supported by the National Science Foundation EAR 1417148 as well as NSF Graduate Research Fellowship Program under Grant No. DGE 1322106. We are partially supported through a DoD contract under the Laboratory of Telecommunication Sciences Partnership with the University of Maryland. We would also like to thank the Maryland Innovation Initiative for their support. We are grateful to Anthony Mautino and John Rzasa for their assistance. We gratefully acknowledge UMD and the office of the Vice President for Research.

## REFERENCES

[1] D. Lathrop, I. Shani, P. Megson, A. Restelli, and A. R. Mautino, "Integrated circuit designs for reservoir computing and machine learning," WO2018213399A1, 2018.

[2] H. Jaeger, "The 'echo state' approach to analysing and training recurrent neural networks-with an Erratum note 1," Bonn, Germany, 2010.

[3] W. Maass, T. Natschläger, and H. Markram, "Real-time computing without stable states: a new framework for neural computation based on perturbations.," *Neural Comput.*, vol. 14, no. 11, pp. 2531–60, 2002.

[4] I. Shani *et al.*, "Dynamics of analog logic-gate networks for machine learning," *Chaos*, vol. 29, 2019. (forthcoming)

[5] Z. Lu, B. R. Hunt, and E. Ott, "Attractor reconstruction by machine learning," *Chaos An Interdiscip. J. Nonlinear Sci.*, vol. 28, no. 6, 2018.

[6] F. Triefenbach, A. Jalalvand, B. Schrauwen, and J.-P. Martens, "Phoneme Recognition with Large Hierarchical Reservoirs," in *Advances in neural information processing systems*, 2010, pp. 2307–2317.

[7] P. Buteneers, B. Schrauwen, D. Verstraeten, and D. Stroobandt, "Epileptic seizure detection using Reservoir Computing," in *Proceedings of the 19th Annual Workshop on Circuits, Systems and Signal Processing*, 2008.

[8] T. Waegeman, F. Wyffels, and B. Schrauwen, "Feedback control by online learning an inverse model Feedback Control by Online Learning an Inverse Model," *IEEE Trans. NEURAL NETWORKS Learn. Syst.*, vol. 23, no. 10, p. 1, 2012.

[9] M. Lukoševičius, H. Jaeger, and B. Schrauwen, "Reservoir Computing Trends," *KI - Künstliche Intelligenz*, 2012.

[10] K. Vandoorne *et al.*, "ARTICLE Experimental demonstration of reservoir computing on a silicon photonics chip," *Nat. Commun.*, vol. 5, 2014.

[11] D. Canaday, A. Griffith, and D. J. Gauthier, "Rapid Time Series Prediction with a Hardware-Based Reservoir Computer," *Chaos An Interdiscip. J. Nonlinear Sci.*, vol. 28, no. 12, p. 123119, 2018.

[12] R. Zhang *et al.*, "Boolean chaos," *Phys. Rev. E - Stat. Nonlinear, Soft Matter Phys.*, vol. 80, no. 4, pp. 1–4, 2009.

[13] D. P. Rosin, D. Rontani, and D. J. Gauthier, "Ultrafast physical generation of random numbers using hybrid Boolean networks," *Phys. Rev. E*, vol. 87, no. 4, p. 040902, 2013.

[14] "PYNQ: Python Productivity for ZYNQ." [Online]. Available: http://www.pynq.io/.